\documentstyle [12pt,twoside]{article}
  
\begin{document}
\begin{flushright}
UBCTP--94--011
\end{flushright}
\centerline{\large{\bf The Interstellar Medium as a}}
\centerline{\large{\bf Gravity Wave Detector}}
\vspace*{1.cm}
\centerline{\bf Redouane Fakir}
\vspace*{0.5cm}
\centerline{\em Cosmology Group, Department of Physics}
\centerline{\em University of British Columbia}
\centerline{\em 6224 Agriculture Road, Vancouver, B.C. V6T 1Z1, Canada}
\centerline{\em fakir@physics.ubc.ca}
\vspace*{1.5cm}
\centerline{\bf Abstract}
\vspace*{0.5cm}

An observer, situated several thousand light-years away
from a radio pulsar, finds himself embedded in the diffraction
pattern resulting from the propagation of the radio waves
through the irregular interstellar medium. The observer's
movement relative to the pattern causes an apparent scintillation
of the pulsar. A binary star, situated close to the pulsar's
line-of-sight, is generating relatively strong gravity waves.

The rays originating from the pulsar experience
a tiny periodic deflection due to the gravity waves produced by
the binary star. This deflection displaces the diffraction
pattern laterally in a manner familiar from refractive interstellar
scintillation, except that this gravity wave effect is not dispersive.
The displacement has the same period as the gravity waves.
Its amplitude equals the product of the tiny deflection angle and
the large distance from the binary star to the observer. This periodic
displacement can reach a few hundred kilometers, which can be
comparable to the size of the features in the diffractive pattern.
Thus, there seems to be a possibility that the exceedingly faint
gravity waves can manifest themselves macroscopically.

Observationally, the end effect could be a substantial, deterministic
alteration of the scintillation time structure.

\clearpage

The importance for fundamental physics and for astrophysics of
an eventual detection of gravity waves is matched only by the
technical difficulty of achieving such a detection [1].

Most effects through which gravity waves can be detected,
such as fractional proper-distance variations, rates of
change in time delays or spatial geodesic deflections,
have magnitudes of the order of $h(D)=H/D$,
where $D$ is the distance from the observer to the source,
and $H$ is a measure of the intrinsic strength of the emission.
Thus, the extreme faintness of the waves is due as much to the
inherent weakness of the emission
as to the largeness of the distances involved
($H << 1m$ and $D\sim 10^{18}$ to $10^{20}m$
for most galactic sources.)

For example, candidate neutron stars have $h(D)\sim 10^{-24}$
in the most optimistic estimates ($D>10^{18}m$ and $H<10^{-6}m$.)
Binary stars ($D>10^{17}m$ and $H<10^{-3}m$) could reach
in principle $h(D)\sim 10^{-20}$, but they emit at much lower
frequencies. This makes their detection by ground-based hardware
extremely difficult, due chiefly to seismic noise [2,3].

An example of high-frequency, intrinsically strong sources is the
merger of two compact stars. Here $H$ should reach about $100m$,
but the expected rareness of such events puts most of them at
extragalactic distances ($D>10^{23}m$.) Still, the resulting
amplitudes from the closest events ($h(D)\sim 10{-21}$) are at
just about the projected threshold of sensitivity of the
LIGO detector, which is already at an advanced stage of development [1,2].

Recently, an alternative approach to gravity-wave detection was
suggested [4-7] that tries to circumvent the stringent
constraint: {\it effect}$\sim h(D)=H/D.$  This was attempted
by considering effects suffered by electromagnetic rays when
they propagate passet a source of gravity waves. It was shown
that if a source of gravity waves lies at a small distance (``impact
parameter'') $b$  from the line-of-sight of an electromagnetic
source, the latter will have its rays deflected by an angle
$\alpha_{gw}\sim\pi^{2}h(b),$ and delayed by an amount $\delta t$ that changes
at a rate $d\delta t/dt\sim h(b)$.

It was then pointed out that there exist several actual astronomical
configurations where one has $b << D$, and hence $h(b) >> h(D)$ for
the same gravity-wave source. A good illustration of this is when a pulsar
is the electromagnetic source, and a neutron star which is
in tight orbit around
it is the gravity-wave source. For instance, for the binary pulsar PSR
B1913+16, $b$ comes to only about $10^{8}m$ every $7^{h}45'.$
For the optimistic estimate $H \approx 10^{-6}m$,
the expected magnitudes of ray deflection and time-delay variation
are $\alpha_{gw}\sim 10^{-7}$'' and $d\delta t/dt\sim 10^{-14} sec/sec.$
This is not too far below today's achievable sensitivities.

These same waves have an amplitude of only $h(D)\sim 10^{-26}$ when they
eventually reach the Earth ($D \approx 1.5\times 10^{20}m$). There is little
hope, in the foreseeable future, of detecting so weak a gravitational
radiation by interfering with it at the level of the solar system.
Thus, the approach proposed in [4-7] amounts to using
electromagnetic waves to probe gravity waves in the vicinity of their
source, where they are still relatively strong.

Another type of astronomical configuration where the above applies
is when the source of gravity waves is a binary star that lies close
to the line-of-sight of the pulsar. For example, the $\mu$-Sco system
has $H\approx 0.06m$ and $D\approx 3\times 10^{18}m$.
Hence, the magnitude of these waves, when they reach the Earth, is only
$h(D)\sim 10^{-20}$.
 If this system were
lying at $1"$ from some distant light source, it would induce
light deflections and time delays comparable to the (very) relatively
large values found above
for the neutron-star case. Furthermore, we shall see, in the following,
that the alignment constraint is less stringent for the effect
described here than for [6,7].

In a sense, then, in the cases identified in [6,7]
the distance $D$ does not suppress
the effect of gravity waves ($effect\sim h(b)=H/b,$)
except perhaps for the deflection effect, which it can sometimes
prevent from translating into angular-position shifts through
the lens equation.

The present letter suggests a way of taking this a step further:
the very largeness of $D$ may contribute
to making gravity waves more accessible to observation.
This could be achieved by exploiting the scattering of
pulsar radio waves by the interstellar medium [8-13].

In what follows, we describe the interstellar scintillation of pulsars
in heuristic terms that summarize an extensive
body of observations and theoretical derivations, about which there
now exist some excellent books and review articles [14-20].

The scattering of pulsar radio waves by inhomogeneities of size $a$
in the interstellar electron density result in a ``diffraction'' pattern
for an observer in the Fresnel region, i.e., at a distance $D>a/\theta_{s}$,
where $\theta_{s}$ is the scattering angle.
The scattering is strong or weak according to whether or not the
phase fluctuation ($\Delta\phi\approx 2\pi\theta_{s}a/\lambda$,
$\lambda$ being the radio wavelength) is larger than about one radian.
Hence, the scattering is strong and the fractional intensity fluctuations
are of order one if the observer is in the Franhofer region:
$D>a^{2}/\lambda$. The velocity ($V$) of the observer relative to the
interference pattern causes the pulsar intensity to appear to scintillate on a
time scale $\tau=S/V$, where $S$ is the typical size of the features
in the pattern. In principle, the effect of internal motions  in the
interstellar medium should be added to that of the relative motions
of the pulsar and the Earth.

Observationally, most pulsars show strong scintillation at wavelengths
$\lambda>10cm$. Observations have also allowed  turbulent
motions within the interstellar medium to be ruled out for most (but not all)
lines-of-sight of interest.

Over the last thirty years or so, the theory of interstellar scintillation
has been developed to the point of being able to account for most of the
phenomena displayed by pulsar radio spectra. However, until relatively
recently, observed long time-scale variations in some
spectral characteristics remained largely unexplained.  Eventually, these
slow variations were shown to be also due to the action of the interstellar
medium, rather than to intrinsic properties of pulsars [12,13].

Interstellar inhomogeneities with scales $a>\theta_{s}D$ do not
contribute to the diffractive scintillation discussed above. Rather,
they cause an overall tilt of the electromagnetic wavefronts,
which are otherwise
randomly perturbed on the smaller diffractive scale. This phenomenon
amounts to a large-scale refraction, which consists in ray bundles being
deflected by an angle $\theta_{r}$. The variation of $\theta_{r}$
on very large scales (refractive focusing), again combined with the
observer's relative velocity, causes deep modulations in the pulsar
intensity over time scales that sometimes exceed several months.

Closer to the subject of the present paper,
refractive scintillation is also responsible for the
frequency drifting of diffractive features which is clearly
present in many dynamic spectra. (These are profiles of intensity versus
radio frequency and time $t$ or spatial position $x=Vt$.)
This is a consequence of the dispersive nature of the refraction by the
interstellar medium, more precisely, of the fact that
$\theta_{r}\propto\lambda^{2}$.

Now, it is a familiar phenomenon in optics (see e.g. [21])
that changing the direction of a
bundle of rays that is incident on an irregular medium by an amount
$\theta_{r}$, shifts the whole pattern laterally by an amount
$\delta s_{r}\approx D\theta_{r}$. (We come back to this more explicitly
below.)
Hence, in the case of refraction by the interstellar medium, for which
$\theta_{r}\propto\lambda^{2}$,
diffractive features at different wavelengths develop a relative shift,
resulting in the kind of frequency drift that has been observed
in many dynamic spectra. The formula describing the drift rate follows
directly from the above by writing
\[
{dt\over d\lambda}  = {dt\over ds} {ds\over d\theta_{r}}
{d\theta_{r}\over d\lambda} = {2 D d\theta_{r}\over \lambda V_{\perp}} \  \  ,
\]
where $V_{\perp}$ is the projection of $V$ unto the direction along
which the refractive ``steering'' takes place.

To the next order of approximation, this steering or group motion
of diffractive features is accompanied by an alteration (usually
slight) of the parameters of the diffractive scintillation.
A detailed study of these refractive alterations can be found in [18].
There, it is shown that the theory of refractive scintillation,
besides being compatible with most observations, is also backed by numerical
simulations.

In the astronomical configurations that we considered earlier (where the
gravity-wave source is close to the line-of-sight of the pulsar)
 the gravity
waves effectively imprint a kink $\alpha_{gw}$ unto the direction of
propagation
of the narrow ray bundle which eventually reaches the vicinity of the
observer. Hence, one has a gravity-wave induced refraction with
parameters $D=D_{GE}$ and $\theta_{r}=\theta_{gw}$. Here $D_{GE}$ is the
distance from the gravity-wave source to the Earth, and
$\theta_{gw}$ is the refraction angle corresponding
to the deflection angle $\alpha_{gw}$. The two
are related through the lens equation:
\[
\theta_{gw} \approx {D_{PG}\over D_{PE}} \alpha_{gw} \  \ ,
\]
where $D_{PG}$ is the distance from the pulsar to the gravity-wave source,
and $D_{PE}$ the distance from the pulsar to the Earth.

The initial narrowness of the ray bundle means that, generically,
the gravity waves induce no significant phase gradient amongst the
relevant rays. (Some possible exceptions will be mentioned towards the
end of this letter.) Also,
the action of the gravity waves decreases rapidly away from
the binary system. Hence, it is negligible during virtually all  the
propagation that
determines
the diffractive scintillation. Thus, the latter is practically the same as in
ordinary
scintillation.
 In all, gravity waves generically leave diffractive
scintillation parameters unchanged (apart from the quasi-rigid shift of the
diffraction pattern) because the gradient in the refraction
angle $\theta_{r}=\theta_{gw}$ (the amount of focusing) is usually
negligible.

In the context
of the present study, the combination of the lens equation and the
diffraction-pattern displacement equation
favors strongly the case where the gravity-wave source is a binary star
placed midway between the pulsar and the Earth , over the case where it is a
neutron star in orbit around the pulsar. This is because, in order to
maximize the pattern displacement $\theta_{gw} D_{GE}$, one would like
to maximize both $D_{PG}=D_{PE}-D_{GE}$ and $D_{GE}$. Thence, the optimal case
is $D_{GE}\sim D_{PE}/2$.

Before moving on, a word about some of the simplifications made here. First,
we have tried to keep the discussion
to the simplest by considering that the observer is confined to a
one-dimensional screen, as it is often done in the interstellar
scintillation literature. Considerations arising from adding more
dimensions [18] are largely irrelevant in the present context,
with the exception of the following aspect.
Besides being non-dispersive, there is another way in
which gravity-wave refraction is distinct from its interstellar
counterpart. When the two polarizations of gravitational radiation
are taken into account, a given feature in the diffraction pattern
will not be oscillating along an exactly straight line, but rather
along a quasi rectilinear segment of a large ellipse centered
around the position corresponding to the absence of gravity waves.
But then, just as in ordinary refractive scintillation, the observable
part of the effect described here is proportional to the cosine
of the angle between the (practically rectilinear) shift $\delta s_{gw}$
and the observer's velocity $V$, i.e., it is proportional to $V_{\perp}$.

This is so only because we are considering a ray bundle that is still
very thin when it propagates passet the gravity-wave source
(thickness much smaller than one gravitational wavelength) and
because then, there is no appreciable differential of gravity-wave phase,
amplitude or polarization over a cross section of the bundle.
At much larger scales than are relevant here, the diffraction pattern
is deformed in a deterministic, albeit non-trivial fashion, due to the
extended nature of the bundle cross-section then considered.
These deformations are closely related to the higher order
gravity-wave effects mentioned at the end of this paper.

Finally, we also do not discuss the effect of the non-radiative
component of the gravitational field generated by the gravity-wave
source. That component produces a deflection that is much larger
than the radiative one, but it follows a very different (and precisely
predictable) time dependence, as the relative motions of
the pulsar, the binary star and the Earth steadily change
the impact parameter $b$. Hence, it should be possible
to subtract that effect, along with that of the
long time-scale refraction that is due to very large-scale interstellar
inhomogeneities.

To see more explicitly how gravity-wave refractive scintillation
behaves, let us write the slowly varying part of the complex scalar
electromagnetic field, first in the absence of refraction,
as (here we follow closely [18]; see also [21])
\newline\begin{equation}
 E(x) = I_{0}^{1/2} \int dx' K_{0}(x-x') e^{i\phi_{d}(x')} \  \ ,
\end{equation}\newline
where the function $K$ is the wave propagator:
\newline\begin{equation}
K_{0}(x) = r_{F0}^{-1} \exp\left\{ i\pi(x/r_{F0})^{2} \right\} \  \ .
\end{equation}\newline
$r_{F0}$ is the Fresnel radius  $r_{F0}=(D_{PE}\lambda)^{1/2}$.
$I_{0}$ is the intensity
of the incident plane wave. $x$ is the spatial coordinate of, say, an
intensity peak on  the
one-dimensional screen. $\phi_{d}$ is the so-called diffractive
phase, generated by the random, statistically homogeneous distribution
of small-scale interstellar fluctuations ($a<D_{PE}\theta_{s}$).

When gravity-wave (or any other) refraction is present,
the phase is incremented by an amount
$\phi_{gw}(x)$ which one can represent by a Taylor expansion
about a reference point $x_{0}$ on the screen:
\newline\begin{equation}
\phi_{gw}(x) = \phi_{0}(x_{0}) + \phi_{1}(x-x_{0})
+ \phi_{2}(x-x_{0})^{2} + ...
\end{equation}

The intensity $I_{0}$ is then multiplied by a factor (``gain function")
$G = (1 + D_{GE}\lambda\phi_{2})^{-1}$ for $D_{GE}\lambda\phi_{2}>-1$
(the speed of light
is set to one throughout.) The pulsar's apparent angular position
shifts by an amount $\theta_{gw}=(x_{gw}-x)/D_{GE}$, where $x_{gw}$ is the
new  position of the peak on the screen, i.e. $x_{gw}=x+\delta x_{gw}$,
where $\delta x_{gw}=\theta_{gw}D_{GE}$ is the quasi-rigid shift mentioned
earlier. One has
\newline\begin{equation}
x_{gw} = G(x-x_{0}) + x_{0} - D_{GE}\lambda\phi_{1} \  \ .
\end{equation}\newline

The electromagnetic amplitude becomes
\newline\begin{equation}
E_{gw}(x)  = (GI_{0})^{1/2} e^{i\psi}
\int dx' K_{gw}(x_{gw}-x') e^{i\phi_{d}(x')} \  \ ,
\end{equation}\newline
where now
\newline\begin{equation}
K_{gw}(x) = r_{F}^{-1} \exp\left\{ i\pi(x/r_{F})^{2} \right\} \  \ ,
\end{equation}\newline
with $r_{F}=G^{1/2}r_{F0}$. ($\psi$ is an irrelevant phase.)

For gravity waves, one has generically $|D_{GE}\lambda\phi_{2}|<<1$,
(see below) so that $G$ is very nearly equal to one.
However, already at the ``zeroth-order'' level ($G=1$), a potentially large
effect takes place: From eqs.(1-6) and $G\approx 1$ we have,
\newline\begin{equation}
E_{gw}(x+\theta_{gw}D_{GE}) \approx E(x) \  \ ,
\end{equation}\newline
indicating that the pattern has shifted quasi-rigidly along
the screen by $\delta x_{gw}\approx\theta_{gw}D_{GE}$, which could
possibly measure in kilometers.

A measure of the observational significance of this effect is the
ratio $\delta x_{gw}/S$, where, as before, $S$ is the typical size of
diffractive features in the pattern.
First, consider an optimal case where a system like $\mu$-Sco, with
$H\approx 0.1 m$ and a gravitational wavelength $\Lambda_{gw}=10^{14}m$,
lies at about $1"$ from the
line-of-sight of a pulsar. That corresponds to $b\approx 10^{14}m$ if the
pulsar is at several kiloparsecs (a typical pulsar
distance), and $D_{GE} \approx 10^{20}m$. This would imply that
$\theta_{gw}^{optimal}\sim 10^{-13}${\it radian}.
Then, the diffractive pattern
could be shifting laterally by as much as
$\delta x_{gw}^{optimal}\approx 10.000 km$.
If one relaxes the alignment constraint from $1"$ to $1'$, the displacement
could still be of the order of $100km$,
a perhaps surprisingly macroscopic manifestation of the exceedingly
faint gravity waves.

The spatial scale $S$, on the other hand, is smaller for longer radio
wavelengths, since
\[
S = {\lambda\over 2\pi\theta_{s}} =
{(a/D)^{1/2}\over r_{e}\Delta n_{e} \lambda} \ \ ,
\]
where $r_{e}$ is the classical electron radius and $\Delta n_{e}$
is the electron density fluctuation for inhomogeneities of size $a$.

Observations yield typically $\theta_{s}\sim 0.2"$ at
$\lambda\sim 3m$, suggesting that $S\sim 1000 km$ at that
wavelength. In practice, $\lambda$ can be increased to about $15m$,
corresponding to $S\sim 200km$, which is much smaller than
$\delta x_{gw}^{optimal}$. In all, even if the gravity-wave source
is either much weaker or much farther from the line-of-sight
than in the optimal case, the
gravity-wave induced periodic shift $\delta x_{gw}$ could still
be of the same order-of-magnitude as the diffraction-pattern
spatial scale $S$.

This concludes our discussion of the lowest-order, and
observationally the most promising of the effects produced by
the gravity waves pointed out in [6,7].

As pointed out earlier, gravity waves may affect the interstellar
scintillation at a higher order if it can induce a significant
phase gradient within the relevant ray bundle.

Gravity waves alone cannot alter significantly the
propagation of electromagnetic waves. Effects such as refractive
focusing were shown by several authors to be unobservable in the
near future [22].
However, when combined with the action of the interstellar
medium, the effect of gravity waves could be made relatively
significant in some cases, although not to the extent of the
lowest-order effect studied above [23].

The physics of these higher-order effects is quite involved,
and the technical details are discussed in a separate publication [23].
Here, we briefly outline the motivation for pursuing that direction
of research.

There are two things that a medium has to do to cause scintillation:
(1) it must bend the rays enough to make them interfere; (2) it must
induce a substantial phase gradient amongst the rays.
Gravity waves are too weak to fulfill the first condition
by themselves. However,  they can, in some cases, fulfill the second
condition. This can be seen by comparing the modulation that they could
induce in the optical-path length ($\Delta L_{gw}$), first with
the radio wavelength $\lambda$, and second with the modulation due
to the interstellar medium ($\Delta L_{ism}$.) It can be shown that
condition (2) could be satisfied at observation wavelengths of a few
centimeters. This is due mainly to the high-power wavelength dependence
of $\Delta L_{ism}$, which is proportional to $\lambda^{4}$.

The hope in this picture is, of course, to make the {\em interstellar medium}
fulfill the condition (1), which gravity waves cannot do. This is, however, a
matter of delicate
balance, since at radio waves as short as a few centimeters, the scattering
power of the interstellar medium itself is considerably weaker
(recall that $\theta_{s}\propto \lambda^{2}$.)

Thus, the question of whether or not the gravity waves, when combined with
the interstellar matter,  can have an observable effect on
scintillation parameters such as the intensity-correlation
frequency bandwidth,
the pulse broadening or the angular broadening, must be decided
by detailed calculation.  This involves computing the gravity-wave
correction to the two-point correlation function of the field
(which gives the angular broadening), the correction to the two-point {\em
frequency}
correlation function (which gives the pulse broadening), and the correction to
the four-point
correlation function (which gives the intensity-correlation
frequency bandwidth.)

One of the difficulties that arise is that one must do the calculations in the
case of {\em weak} interstellar scintillation, as we said. Even in
the absence of gravity waves, analytical approximations to the
four-point correlation function were found only in the case of
strong scintillation, where it can be expressed in terms of lower-order
moments.

What we have just said perhaps sheds some light on why previous
researchers, who looked for observable effects of gravity waves
on electromagnetic propagation [22], or for gravitational effects
on interstellar scintillation [24], concluded (correctly) to the negative.
In [22], the inhomogeneous interstellar matter is {\em replaced}
by (rather than added to) a background of gravity waves. Then, as
we said, condition (1) is hard to fulfill. In [24], the interstellar
matter is replaced by  the stochastic sum of the non-radiative gravitational
fields of the stars that lie along the line-of-sight of a given pulsar.
In this case, condition (1) may be fulfilled, but generally not condition (2):
The variation in the optical path induced by the Schwarzschild
metric decreases very slowly with the impact parameter [25]: it varies
essentially as $\log(b)$. This is in contrast with the gravity-wave
cases of [6,7], where that dependence is close to $1/b$. This implies
that, in [24], the phase differential induced in a typical ray bundle
tends to be small, hence condition (2) is unlikely to be satisfied.

In conclusion, there seems to be a possibility that
the extremely tenuous gravity waves can manifest themselves
macroscopically. If true, much experimental work, in terms of noise
reduction, would remain to be done before such manifestations
can be clearly read from the data. In particular, noise reduction
by exploitation of the high regularity of the waves should be more
efficient for shorter-period binaries (periods range all the way down
to days or even hours.) Fortunately, there has been recently a flurry
of activity in the field of high-resolution astrometry [26,27],
which should greatly enhance the chances of finding conveniently
located binary stars. Another direction of research would be to look
for the much more scarce cases where the conveniently located gravity-wave
source emits at a much
higher frequency than considered here. These cases could involve
neutron stars or gravitational
pulses from violent astrophysical phenomena. In both cases, the frequencies
could reach $100Hz$ or more [23]. The effect on the interstellar scintillation
would then take on a rather different apparence than here, since the velocity
of the diffraction pattern  due to gravity waves could then exceed
the velocity of the observer.

\vspace*{2.cm}
\centerline{\bf Acknowledgements}
\vspace*{0.5cm}

I have greatly benefited from several discussions with W.G. Unruh.
I am also grateful to J.M. Cordes and R.V.E. Lovelace for helping
with understanding interstellar scintillation, and to
W.L.H. Shuter, J. Meyer and W.F. Dalby for some stimulating conversations.
I am equally thankful to T.E. Vassar for helping with the
manuscript.

This work was made possible by an
extended logistical support from the Cosmology Group in the
Department of Physics, University of British Columbia.
\clearpage
\centerline{\bf References}
\vspace*{1.cm}
[1]  K.S. Thorne, in {\em 300 Years of Gravitation}, S.W.
Hawking and W. Israel, Eds.(Cambridge University Press, Cambridge, 1987.)
\newline\newline
[2]  R.E. Vogt, {\em The U.S. LIGO Project} in
{\em Proc. of the Sixth Marcel Grossmann Meeting on GRG, MGC, Kyoto,
Japan, 1991.}
\newline\newline
[3]  C. Bradaschia et. al., {\em Nucl. Instrum. \& Methods},
A289, 518 (1990).
\newline\newline
[4]  R. Fakir (1991), {\em Gravity Waves and Hipparcos},
proposal presented in {\em Table Ronde on Moroccan-European Collaboration
in Astronomy, Moroccan Center for Scientific Research Archives,
June 1991, Rabat,
Morocco.}
\newline\newline
[5]  R. Fakir (1993), {\em The Astrophysical Journal, vol.418, 202}.
\newline\newline
[6]  R. Fakir (1994), {\em The Astrophysical Journal, vol.426, 74}.
\newline\newline
[7]  R. Fakir (1994), {\em Physical Review D, vol.50, 3795}.
\newline\newline
[8] A.G. Lyne and B.J. Rickett (1968), {\em Nature, vol.218, 326}.
\newline\newline
[9] P.A.G. Scheuer (1968), {\em Nature, vol.218, 920}.
\newline\newline
[10] B.J. Rickett (1969), {\em Nature, vol.221, 158}.
\newline\newline
[11] E.E. Salpeter (1969), {\em Nature, vol.221, 31}.
\newline\newline
[12] W. Sieber (1982), {\em Astron.Astrophys., vol.113, 311}.
\newline\newline
[13] B.J. Rickett, Wm. A. Coles and G. Bourgois (1984),
           {\em Astron.Astrophys., vol.134, 390}.
\newline\newline
[14]  A. Hewish (1975), {\em Science, vol.188, 1079}.
\newline\newline
[15]  B.J. Rickett (1977), {\em Ann.Rev.Astron.Astrophys., vol.15, 479}.
\newline\newline
[16] B.J. Uscinski (1977), {\em ``The elements of wave propagation in
            random media'' (pub.: McGraw-Hill.)}
\newline\newline
[17] R.N. Manchester and J.H. Taylor (1977),
       {\em ``Pulsars'' (pub.: Freeman and company, San Fransisco.)}
\newline\newline
[18] J.M. Cordes, A. Pidwerbetsky and Lovelace (1986),
        {\em The Astrophysical Journal, vol.310, 737}.
\newline\newline
[19] A.G. Lyne and F. Graham-Smith (1990),
       {\em ``Pulsar astronomy'' (pub.: Cambridge University Press.)}
\newline\newline
[20] B.J. Rickett (1993), {\em in ``Wave Propagation in Random Media''
       (eds.: Tatarski, Ishimaru and Zavorotny; pubs: IOP and SPIE press.)}
\newline\newline
[21] M. Born and E. Wolf (1959), {\em ``Principles of optics''
          (pub.: Pergamon Press.)}
\newline\newline
[22]  D. Zipoy (1966), {\em Phys.Rev., vol.142, 826}.
See also P.G. Bergmann (1971), {\em Phys.Rev.Lett., vol. 26, 1398},
P.J. Adams, R.W. Hellings and R.L. Zimmermann (1984),
{\em Astrophys. J. Lett., vol. 280, L39}, and
Braginsky, Kardashev, Polnarev, and Novikov (1990), {\em Nuovo
Cimento, vol.105, 1141}.
\newline\newline
[23]  R. Fakir (1994), {\em Gravity Waves and Light},
in preparation.
\newline\newline
[24]  L.M. Erukhimov and P.I. Shpiro (1993), {\em in ``Wave Propagation in
Random Media''
       (eds.: Tatarski, Ishimaru and Zavorotny; pubs:IOP and SPIE.)}
\newline\newline
[25]  I.I. Shapiro (1964), {\em Phys.Rev.Lett., vol.13, 789}.
D.O. Muhleman  and P. Reichley (1965),
{\em JLP space programs summary 4, no.37-31, 239}.
\newline\newline
[26]  M.A.C. Perryman et al (1992),
{\em Astronomy \& Astrophysics,vol.258,1}.
\newline\newline
[27]  E. H\o g (1993), in
{\em ``Developments in Astronomy and Their Impact on
Astrophysics and Geodynamics'' (eds: Mueller and Kolaczek.)}

\end{document}